\documentclass[12pt]{JHEP3}
\usepackage{amsfonts}
\usepackage{epsfig}
\def\ss{\scriptscriptstyle}

\def\pl{\partial}

\def\Dsl{\hbox{/\kern-.6700em\it D}} 
\def\dsl{\hbox{/\kern-.5300em$\partial$}}

\def\eqa{\begin{eqnarray}}
\def\eeqa{\end{eqnarray}}
\def\eq{\begin{equation}}
\def\eeq{\end{equation}}

\def\pref#1{(\ref{#1})}
\def\mui{\mu_{\ss I}}

\title{Instabilities and Particle Production in S-Brane Geometries}
\author{C.P. Burgess$^1$, P. Martineau$^1$, F. Quevedo $^2$,
 G. Tasinato $^3$ and I. Zavala C.$^2$
\\

$^1$ Physics Department, McGill University,  3600 University Street,
 Montr\'eal, Qu\'ebec, Canada, H3A 2T8.\\

$^2$ Centre for Mathematical Sciences, DAMTP,
               University of Cambridge,\\
               Cambridge CB3 0WA UK.\\

$^3$ Physikalisches Institut der Universit\"at Bonn, \\
     Nussallee 12, 53115 Bonn, Germany
}

\abstract{ We study the classical stability of a class
of S-brane geometries having cosmological horizons. By considering
the perturbations of the metric in these geometries we establish
that their horizons are unstable in the sense that an observer
trying to cross the horizon experiences an infinite flux of
radiation at the instant of crossing. The backreaction of this
radiation is likely to convert the horizons into curvature
singularities, similar to the instability of the internal Cauchy horizon
of the Reissner-Nordstr\"om black hole. We also compute the
particle production by the time-dependent fields in the future
regions of these geometries, and find that the spectrum of
produced particles is thermal, with
temperature coinciding with  the Hawking temperature computed
by euclideanizing the metric in the static region. Possible
implications of these results are discussed.  \vskip0.4cm \hfill\break
\leftline{DAMTP-2003-5, hep-th/0301122} } 

 \preprint{}

\def\eq{Eq.\,}

\def\be{\begin{equation}}
\def\ee{\end{equation}}
\def\bea{\begin{eqnarray}}
\def\eea{\end{eqnarray}}

\begin{document}

\section{Introduction}
S-branes are spacelike surfaces in spacetime along which
transitions between different vacua are speculated to take place
in string theory \cite{sbranes}. Their study is motivated by the
desire to be able to extend the present theoretical toolbag to
include techniques for understanding time-dependent problems in
string theory. If the transitions described by S-branes really do
arise, parts of their behavior should be describable in terms of
the evolution of the low-energy fields of supergravity, perhaps
also with a rolling tachyon or tachyons which describe the
field-theory version of the transition between vacua.

Several time-dependent supergravity solutions have been proposed
as corresponding to S-branes
\cite{oldsolutions,gqtz,newsolutions,rob,bqrtz}. Although many
have an FRW-like singularity at early times, some do not and
instead have an interesting global structure consisting of
non-singular and asymptotically-flat past and future regions
separated by static regions having time-like singularities. Some
of these geometries have been studied in fair detail. In
particular, Ref.~\cite{bqrtz} performs a general analysis of the
charge, tension, entropy and Hawking temperature of many of these
spaces. It was concluded there that the time-dependent regions
provide an interesting interpretation that fits very well with the
original S-brane proposal \cite{sbranes}. The static regions could
be interpreted as the fields external to a pair of
oppositely-charged, negative-tension branes. A similar
interpretation was proposed earlier in terms of orientifold planes
\cite{ckk}.

Our purpose here is to explore two important issues of stability
for these geometries, which were not completely addressed in
Ref.~\cite{bqrtz}. The first of these is the classical stability
of the solutions. Our preliminary study showed that small
perturbations in Klein-Gordon scalar fields in these geometries do
not grow with time, but {\em do} infinitely blue-shift for
inertial observers passing through the horizons. This signals a
potential instability, because the backreaction of this energy
density on the metric is ultimately likely to convert the horizons
into curvature  singularities \cite{poisson}.
Here we extend this observation somewhat by performing a similar
analysis for the modes of the metric itself, adapting for this
purpose the methods applied by Chandrasekhar and Hartle to the
Reissner-Nordstr\"om black hole. We arrive at the same conclusion
as before, thereby strengthening the arguments that the horizons
of this geometry are {\it unstable}. This fact indicates that
gravitational perturbation will generally introduce {\it null-like
singularities}   in the geometry.

The presence of these singularities makes these solutions more
similar in character to the general S-brane solutions discussed in
\cite{rob}, for which similar curvature singularities are already
present on null-like surfaces at the classical level. Only for
special values of some parameters do the singular surfaces of
\cite{rob} become horizons, and based on our calculation one might
wonder if these horizons are stable to the formation of
singularities once metric fluctuations are considered.

The second question we address concerns the particle production
which is produced by the non-static metric at late times. In
ref.~\cite{bqrtz} a Hawking temperature was associated with the
static regions of the metric, by requiring in the usual way that
there be no conical singularities in their Euclidean sections.
This temperature was argued to be related to the particle spectrum
which is seen by the (accelerating) static observers. Here we
directly compute the particle production in the non-static region
and show that it is also thermal in character, with the
same temperature that was found previously for the static regions.

\section{Classical Instability} \label{clinst}

In this section we present  in detail the stability analysis for 
gravitational metric
perturbations  of the simplest S0-brane geometry of ref.~\cite{bqrtz}. 
We will closely follow a procedure used
by Chandrasekhar and Hartle \cite{chandra1,chandra}, to show the
instability of the Reissner-Nordstr\"om (RN) metric.
We show that the same conclusion also applies to the S0-brane
geometries of interest here.

\FIGURE{
\let\picnaturalsize=N
\def\picsize{2.0in}
\def\picfilename{stability.eps}
\ifx\nopictures Y\else{\ifx\epsfloaded Y\else\input epsf \fi
\let\epsfloaded=Y
\centerline{\ifx\picnaturalsize N\epsfxsize \picsize\fi
\epsfbox{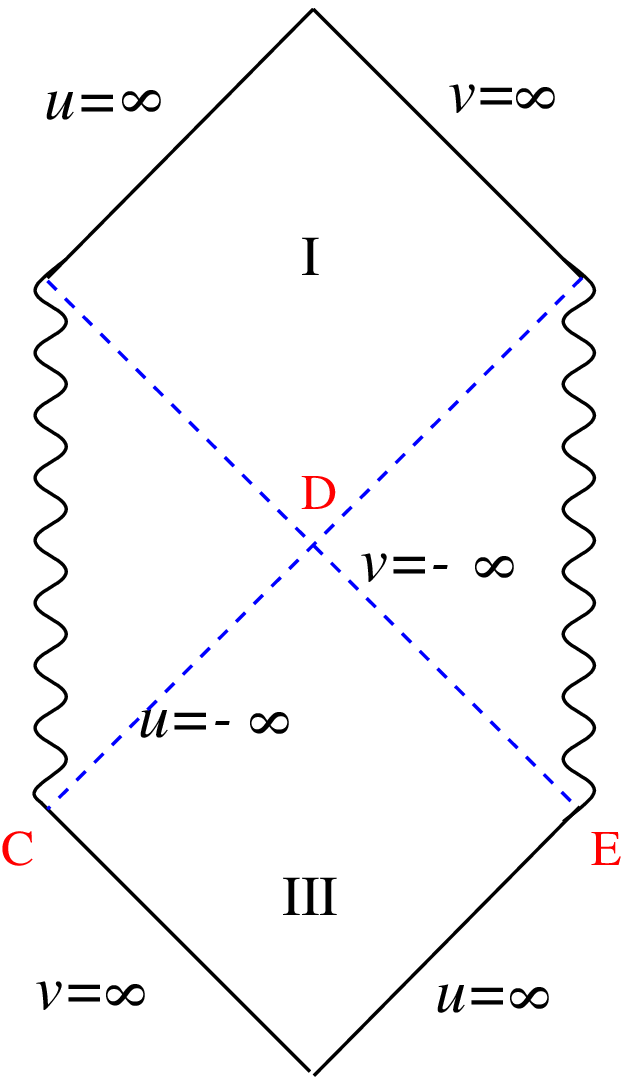}}}\fi
\caption{Penrose diagram for the study of
stability.\label{sbrdraw}} }

We start by reviewing the preliminary analysis performed in
ref.~\cite{bqrtz}, where the Klein-Gordon modes were analysed arriving
to the preliminary conclusion of instability of the Cauchy horizon. We
then perform a more rigorous analysis by considering in detail the metric
perturbation modes, arriving to the same conclusion of instability. 

Consider then the simplest geometry that  corresponds to an 
uncharged S0-brane in $4$ dimensions, whose Penrose diagram is given in Fig.
\ref{sbrdraw}, and is simply a $\pi/2$-rotation of the Penrose
diagram for the Schwarzschild black hole. (The analysis for a
charged brane can be done in analogy with the present case, with virtually no
modification to the equations we present here, and with the same
results.) The metric for this spacetime is:
\be\label{metric}
ds^2\ =\ -\ \left(1-\frac{2p}{t}\right)^{-1}\ {dt^2}\ +\
\left(1-\frac{2p}{t}\right)\ dr^2\ +\ t^2 d{\theta}^2\ +\
t^2\sinh^2{\theta}\ d{\phi}^2\,,
\ee
where
\be\label{ache}
h(t) = 1-\frac{2p}{t}\,.
\ee
The coordinate $t$ is a time coordinate in
regions I and III of Fig.~\ref{sbrdraw}, but is a spatial
coordinate in the other two regions.

The coordinates of eqs.~\pref{metric} and \pref{ache} break down
on the surfaces $t = 2p$, which correspond to the diagonal lines
which form the boundaries of regions I and III. These are the
horizons for this geometry. From the Penrose diagram it is easy to
see that the line CDE is  a Cauchy horizon, since the
initial-value problem in region III does not uniquely determine
the field evolution at points to the future of this line. Field
evolution past the line CDE is not unique because it can also be
influenced by signals from the time-like singularities.

The Klein-Gordon equation for a massive scalar field
propagating in the background eq.(\ref{metric}), is given by 
\bea 
- \; \frac{1}{\sqrt{g}}\partial_{M}
\left[\sqrt{g}g^{MN}\partial_{N}\right] \psi + M^{2}\psi = 0
\nonumber \eea
in the time-dependent regions I and III. Now, since we are interested
in the near horizon limit of the modes, it is convenient to write
the KG equation in terms of isotropic coordinates defined as $\tau =
t-2p$;  the equation is then given by
\bea - \; \frac{1}{\sqrt{g}}\partial_{\tau} \left[ \sqrt{g}
g^{\tau\tau} \partial_{\tau} \right] \psi - \; g^{rr}
\partial_{r}^2 \psi - \; \frac{1}{\tau^2 H_+\,\sqrt{h}} \partial_{i}
\left[\sqrt{h} h^{ij} \partial_{j} \right] \psi + M^{2}\psi = 0 .
 \eea
Here, for clarity, we denote $h_{ij}(\theta,\phi)$ for the metric on the
$2$-dimensional maximally-symmetric hyperbolic space, and write 
$g_{ij}(\tau,\theta,\phi) = \tau^2 H_+ h_{ij}(\theta,\phi)$, where $H_+=
1+ 2p/\tau$. The relevant metric components are:
$g_{\tau\tau} = - H_+$ 
 and $g_{rr} =  H_+^{-1}$. 
%
\noindent The functional form of the metric involved permits 
separation of variables, so we take $\psi(r,\tau,\theta,\phi) =
{\rm e}^{i \sigma\,r} \, f(\tau) \, L_k(\theta,\phi)$, where $\sigma$
and $k$ are separation constants determined by the eigenvalue equations:
\bea - \partial_r^2 e^{i \sigma\, r} = \sigma^2 e^{i \sigma\, r} 
\qquad \hbox{and}
\qquad - \; \frac{1}{\sqrt{h}} \partial_{i} \left[ \sqrt{h} h^{ij}
\partial_{j} \right] L_k = k^2 L_k. \nonumber \eea
Both eigenvalue equations can be solved explicitly, and
delta-function  normalizability of the solutions
require both $\sigma^2 \ge 0$ and $k^2 \ge 0$. The temporal eigenvalue
equation then becomes:
\be \label{KGteq} - \; \frac{1}{\sqrt{g}} \frac{d}{d\tau} \left[
\sqrt{g} g^{\tau\tau} \frac{df }{d\tau} \right] + \left[g^{rr} \sigma^2  +
\frac{k^2}{\tau^2 H_+}  + M^{2}\right] f = 0 . 
\ee
\smallskip
Near the horizon, $\tau \to 0$ and the asymptotic form is governed
by the limits $H_+ \to 2p/\tau$. The metric functions 
therefore reduce to $g_{\tau\tau} \to \alpha_\tau \tau^{-1}$,
$g_{rr} \to \alpha_r \tau$ and $\omega \to \alpha_\omega$.
The precise values of the constants $\alpha_\tau, \alpha_r$ and
$\alpha_\omega$ are not required, apart from the following ratio:
${\alpha_\tau \over \alpha_r}  = r_+^2$. 
With these limits, the Klein-Gordon equation becomes, in the
near-horizon limit:
\be\label{KGthor}
 \ddot f + \frac{1}{\tau} \dot f + \left[ {\alpha_\tau \sigma^2
 \over \alpha_r } {1 \over \tau^2}+
 \alpha_\tau \tau^{-1} \; \left( M^2 + {k^2 \over \alpha_\omega^2 }
 \right) \right] f = 0\,,
\ee
If $\sigma\ne 0$, then the solutions are oscillatory, having the form
$f \sim \tau^{a_0}$, with $a_0 = \pm i \sigma
\sqrt{\alpha_\tau/\alpha_r}$. If $\sigma=0$, then a similar argument
shows that the solutions are nonsingular as $\tau \to 0$.

One can now estimate  whether an instability does
exist by computing the energy, $E = - u^m \partial_m \psi$ of the
Klein-Gordon modes considered above, as seen by an observer whose
velocity, $u = M \partial_t + N \partial_r$, is well-behaved as it
crosses the horizon. The normalization condition $u^2 = -1$ in the
vicinity of the horizon allows a determination of how $M$ and $N$
must behave as $\tau \to 0$ (in isotropic coordinates) in order to
remain non-singular. We find in this way $u^2 \sim -\alpha_\tau M^2
\tau^{-1} + \alpha_r N^2 \tau$, which is regular near $\tau
\to 0$ provided $M \sim \tau^{1/2}$ and $N \sim
\tau^{-1/2}$ near the horizon. With this choice, one then
finds
\bea 
-E = M \partial_\tau \psi + N \partial_r \psi \sim \psi \;
\tau^{ -1/2} \, .  
\eea
Using the asymptotic solution found below eq.(\ref{KGthor}): $\psi
\sim \tau^{a_0}$ with $a_0 = \pm i \sigma \sqrt{\alpha_\tau/\alpha_r}$,
we see that $E \to \infty$ as the horizon is approached.  This
suggests that the stress-energy density of the mode under
consideration diverges as well in this limit. As such, this mode
is likely to destabilize the metric modes near the past horizon.
\smallskip

We will now confirm this result by a full analysis of the metric
perturbations.
For later purposes it is convenient to define several quantities.
We define the surface gravity of the solution to be
\be
\kappa_0 = \frac{1}{2}\Big| \frac{d h(t)}{dt}\Bigg|_{t=2p}\,,
\ee
and  the `tortoise' coordinate as
\be t_* \equiv \int \frac{dt}{h(t)} = t + \frac{1}{2\kappa_0} \ln
|t-2p| \,,
\ee
where $-\infty < t_* < +\infty$ corresponds to the range $2p < t
< \infty$. Notice that $t_*$ increases from past to future in
region I, but decreases from past to future for region III.

Our focus is on metric perturbations in region III of the Penrose
diagram, which is to the past of the Cauchy horizon. We also focus
on the `axial' perturbations of the metric, which are defined as
follows \cite{chandra}:  Take one of the angular coordinates
$\phi$ and change its line element by
\be
d\phi^2\rightarrow\left(d\phi- q_1dr-q_2 dt- q_3 d\theta\right)^2
 \ee
where the $q_i$'s are arbitrary functions of $r,\,t$ and $\theta$.
The metric perturbations can be written, when specialized to the
`axial' modes, as a scalar equation for the field,
$\Phi(t,r,\theta) = t^2 h(t)\,
\left(q_{2,\theta}-q_{3,t}\right)\sinh^3{\theta}$
(see~\cite{chandra}  for details).
Given the symmetries of the problem,  the field equation for $\Phi$
can be  solved by separation of variables, with the field
$\Phi(t,r,\theta)$ decomposed as
\be \Phi(t,r,\theta)=t^{-1} Z(t) \,\Theta_k(\theta)\, e^{i \sigma r} \,,
\ee
with the functions ${\Theta}_k(\theta)$ and $e^{i \sigma r}$, defined
as solutions to the following eigenvalue equations \cite{chandra}:
\begin{eqnarray*}
- \frac{\partial^2}{\partial r^2} \; e^{i\sigma\, r} = \sigma^2
e^{i\sigma\, r}\,, \qquad \mbox{and} \qquad
\sinh^3{\theta} \frac{d}{d\theta} \left(
\frac{1}{\sinh^3{\theta}}\frac{d}{d\theta} \Theta_k \right)
 = k^2 \Theta_k \,,
\end{eqnarray*}

Writing the relevant equations in terms of the tortoise coordinate
defined above, the temporal eigenvalue equation then becomes:
\be\label{perturbation} \frac{d^2 Z(t_*)}{dt^2_*} + \sigma^2
Z(t_*) = V(t(t_*)) Z(t_*)\,, \ee
where the potential in the previous equation is given by
\be \label{Vform}
V(t) = - \frac{(t-2p)}{t^{3}}\left[
\frac{2p}{t} - 2\, h(t) + k^2 \right]\,. \ee

Our interest is in the behavior of the mode $\Phi$ near the
horizon in region III ($t_* \to - \infty$), given its form in the
asymptotic past ($t_* \to \infty$). We therefore need the
asymptotic behavior of the potential in these limits, which is
\be \label{asymptot}
 V(t_*) \propto  \cases{1/t^2 & for $t_* \to t \to \infty$ \cr
e^{2\kappa_0 t_*} & for $t_* \to -\infty$. \cr} \ee
Since the potential falls off faster than $1/t_*$, the asymptotic
behavior of the solutions of (\ref{perturbation}) for $t_* \to \pm
\infty$ is given by $e^{\pm i \sigma t_*}$.

Our initial condition must have only incoming waves in the
asymptotic past, which for region III means for $t \to + \infty$.
Given this we wish to compute the coefficients $A(\sigma)$ and
$B(\sigma)$ which control the behavior of the solutions near the
horizon, according to
\bea \label{bdya}
    Z(t_*) && \to \phantom{A(\sigma) \, e^{-i \sigma t_*} + B(\sigma)\, }
     e^{i\sigma t_*} \qquad
     t_* \to t \to \infty \\
 && \to A(\sigma)\, e^{-i \sigma t_*} + B(\sigma)\, e^{i \sigma t_*}
       \qquad    t_* \to -\infty \,.\label{bdyb}
\eea
(These boundary conditions look slightly odd compared to the Black
Hole case, due to the reversal of roles between $+\infty$ and
$-\infty$ which may be traced to the unconventional property that
$t_*$ becomes more negative into the future for region III (see
Fig.~1). This difference of boundary conditions represents a key
difference between the S0-brane geometry of interest here and
other similar geometries, like the Reissner-Nordstr\"om metric.)

In this way, the problem is reduced to that of scattering of waves on
the potential of eq.~\pref{Vform}. In order to distinguish transmission from
reflection for a given initial wave we must distinguish the edges
$CD$ and $DE$ of the Penrose diagram. In order to do so, we restore the
$r$-dependence, $e^{i\sigma r}$, of the initial wave, corresponding
to a wave whose wavefronts initially move towards increasing $r$.
It is convenient to express the solutions in terms of the
null-like coordinates $u = t_* + r$, $v = t_* -r$, in which case
the initial configuration is $Z(t_*,r) \to e^{i\sigma u}$ for $t
\to \infty$, while the near-horizon limit of eq.~(\ref{bdyb})
becomes
\be\label{forZ} Z(t_*,r) \to e^{i\sigma u} + [B(\sigma)-1] \,
e^{i\sigma u} + A(\sigma) \, e^{-i\sigma v}\,. \ee
From the previous expression (or from looking at the figure) it is
clear that the transmitted part of the wave will cross the edge
$CD$, while the reflected part crosses edge $DE$.

If we take a general, properly weighted,  initial amplitude
 $W(\sigma)$ 
then the above formulae can be rewritten as
\be Z(t_*,r) \to X(v) + Y (u)  \qquad (u,\,v \to -\infty ) \ee
where
\bea\label{x}
    && X(v) = \int_{-\infty}^{\infty}{W(\sigma) \,
    A(\sigma) \, e^{-i\sigma v}d\sigma}\,,\\
    && \label{y} Y(u) = \int_{-\infty}^{\infty}{W(\sigma) \,
    [B(\sigma)-1] \, e^{i\sigma u} d\sigma}\,.
\eea

To argue for instability we now compute the energy contained in
the radiation as seen by a radially-moving inertial observer
crossing the Cauchy horizon. The $(n+2)$  velocity, ${\bf U}$, of such
an observer is given by \cite{bqrtz}:
\be
U^r = \frac{dr}{d\tau}= \frac{E}{h(t)}\,; \qquad \quad
       U^{t_*} = \frac{dt_*}{d\tau}=- \frac{1}{h(t)}[E^2 -
h(t)]^{1/2}\,,  \qquad \quad U^i=0\,,
\ee
where $\tau$ is the proper time and we choose the negative sign
for $U^{t_*}$ in region (III) so that ${\bf U}$ is
future-directed. Notice that the integration constant, $E$, which
labels the observer's geodesic can be negative.

A measure of the energy of the fluctuation $\Phi$ as seen by this
observer is given by ${\mathcal F}$, defined by
\be
\mathcal F = U^\mu Z_{,\mu} = U^rZ_{,r} + U^{t_*}Z_{,t_*}\,,
\ee
or:
\be\label{flujo1}
\mathcal F = \frac{1}{h(t)}\left[ EZ_{,r} - (E^2 - h(t))^{1/2}Z_{,t_*}
          \right]\,.
\ee
In terms of $X$ and $Y$, in the near-horizon limit we have
\bea
&& Z_{,r} \to   \quad   \,\,\,\,\,\, X_{,-v} +  Y_{,u}\,,\\
&&  Z_{,t_*} \to \quad   -X_{,-v} +  Y_{,u}\,,
\eea
and so eq. (\ref{flujo1}) becomes
\be\label{flujo2}
\mathcal F \to \frac{1}{h}\left[ X_{,-v}\left(E+[E^2-h]^{1/2}\right)
           + Y_{,u}\left(E - [E^2 - h]^{1/2}\right)\right]\,.
\ee

We now ask whether ${\cal F}$ diverges on one of the horizons, $CD$ or
$DE$. On $CD$,  $v$  remains finite, while $u\to
-\infty$. This means  that, for $E<0$, the term involving $X_{,v}$ of
(\ref{flujo2}) remains finite while the term with $Y_{,u}$ could
diverge. Hence:
\be\label{flujoCD}
\mathcal F_{CD} \to -4 p\, |E| Y_{,u}e^{-\kappa_o u}\, \qquad
(u\to -\infty \,\,\,\,{\rm on}\,\,\,\,CD)\,. \ee
On $DE$, by contrast, it is $u$ which remains finite, while $v\to
-\infty$. In this case, when $E>0$, the term involving $Y_{,u}$
remains finite but the term with $X_{,-v}$ diverges. Hence
\be\label{flujoDE}
\mathcal F_{DE} \to 4p\, E X_{,-v}e^{-\kappa_o v}\,
       \qquad (v\to -\infty \,\,\,\,{\rm on}\,\,\,\,DE)\,.
\ee

From these expressions we see that the existence of a divergence
in ${\cal F}$ depends on the behavior of
\bea\label{xprima} &&  X_{,-v}= \int_{-\infty}^{\infty}
{W(\sigma)\, i\sigma \, A(\sigma) \,
           e^{-i\sigma v} \; d\sigma}\,,\\
&& \label{yprima}  Y_{,u}= \int_{-\infty}^{\infty} {W(\sigma) \,
i\sigma \,  [B(\sigma)-1] \, e^{i\sigma u} \; d\sigma}\,, \eea
near the horizons. The integrals may be evaluated by contour
integration, with the contour closed in the {\it lower} half-plane
for $Y_{,u}$ (where our interest is in $u\to -\infty$) and in the
{\it upper} half plane for $X_{-v}$ (where we care about $v\to
-\infty$). The result depends on the singularities of $A(\sigma)$
and $B(\sigma)$, which have analytic properties which can be quite
generally determined using the arguments of Chandrasekhar and
Hartle \cite{chandra1,chandra} with virtually no modification. One
finds in this way the function $A(\sigma)$
 is analytic on the upper
half plane except for the poles located at $i n  \kappa_{0}$, where
$n$ is a natural number,  while the  function $B(\sigma)$ is analytic
 on the entire upper half plane.
 This singularity structure is
illustrated in Fig.~2.

\FIGURE{
\let\picnaturalsize=N
\def\picsize{3.0in}
\def\picfilename{analityc.eps}
\ifx\nopictures Y\else{\ifx\epsfloaded Y\else\input epsf \fi
\let\epsfloaded=Y
\centerline{\ifx\picnaturalsize N\epsfxsize \picsize\fi
\epsfbox{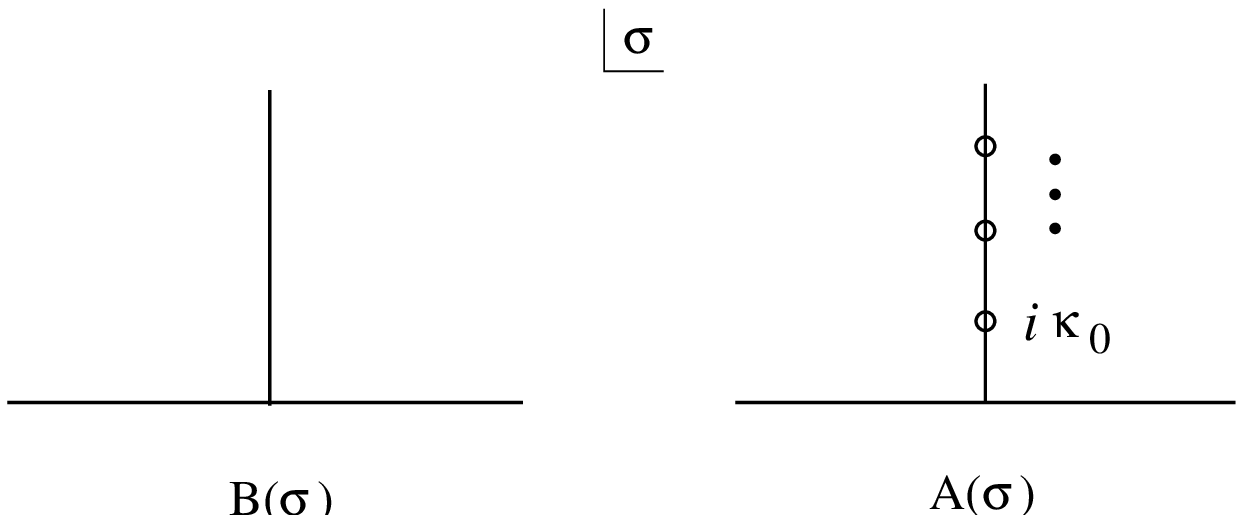}}}\fi
\caption{Domains of analyticity of $A(\sigma)$ and $B(\sigma)$.} }

Applying these results to the flux on surface $DE$, we evaluate
the integral by deforming the contour into the upper half plane.
The dominant contribution is then given by the closest pole to the
real axis, which occurs at $\sigma = i\kappa_0$. Then the integral
giving $X_{,-v}$ becomes
\be X_{,-v} \propto e^{\kappa_0 v}\,. \ee
We see, from (\ref{flujoDE}), that ${\mathcal F_{DE}}$, is
therefore bounded as the observer crosses the horizon.

Similarly evaluating the integral for $Y_{,u}$ by closing the
contour in the lower half-plane, we find contributions only from
the real axis: $i\sigma =0$. Consequently, for $W(\sigma)$
analytic on the real axis, $Y_{,u}$ is $O(1)$ near the horizon
$CD$. It then follows from (\ref{flujoCD}) that the flux $\mathcal
F_{CD}$ at $CD$ necessarily {\it diverges}.

Although we have derived the instability for the lower Cauchy horizons,
we also expect the same conclusion to hold for the upper horizons
which bound region I of the Penrose diagram. We can argue this on
grounds of continuity, as we follow the energy seen by a family of
observers who cross the horizons in the vicinity of point $D$ of
the Penrose diagram. We expect from this that the same infinite
blue-shift seen by inertial observers for horizons $CD$ and $DE$
also extends to the other two horizons.

This calculation confirms the preliminary work of
ref.~\cite{bqrtz}, which argued for infinite energy for
Klein-Gordon modes on this metric. The present calculation shows
that the same conclusion also holds for {\it bona fide} metric
modes. The existence of this divergent energy strongly suggests
that the horizon is unstable towards becoming a curvature
singularity, due to the metric's back-reaction of these large
energy densities: gravitational
perturbations will introduce null-like singularities in the geometry.
This conclusion strongly resembles the same
results for the instability of the RN black hole.

The previous calculation can be extended to more general
asymptotically-flat time-dependent backgrounds representing
cosmological horizons. In general, the temporal eigenvalue
equation that rules the perturbations will take the form of a
Schr\"odinger equation like eq.~(\ref{perturbation}). When the
resulting potential is characterized by an asymptotic behavior
given by (\ref{asymptot}), the resulting cosmological horizon will
be unstable under cosmological gravitational perturbations. Also,
the analysis of `polar' perturbations gives similar results as
does the axial case presented above \cite{chandra}.

\section{Particle Production}
We now turn our attention to  the time-dependent region (region I)
which is to the future of all of the horizons. Because the metric
in this region is not static, it should cause particle production
for any quantum fields which propagate within it. We compute this
particle production here, and show that it  has features which
resemble a thermal distribution whose temperature is given by the
Hawking temperature, as defined in ref.~\cite{bqrtz} for the {\it
static} part of the metric. We find in this way a connection
between the properties of the time-dependent region I and the
static regions which are separated from it by the horizons. These
did not {\it a priori} need to be related, and such a relation
seems even more odd if the surface dividing these regions
represents a curvature singularity rather than a horizon.

To this end we again consider the simplest non-trivial case,
namely the S0-brane in four dimensions, with metric given by
(\ref{metric}),
\be{ds^2\ =\ -\ \left(1-\frac{2p}{t}\right)^{-1}\ {dt^2}\ +\
\left(1-\frac{2p}{t}\right)\ dr^2\ +\ t^2 d{\theta}^2\ +\
t^2\sinh^2{\theta}\ d{\phi}^2}.\ee

\smallskip

Let us first recall the formal calculation of the Hawking
temperature performed in \cite{bqrtz}.  This requires the
cancellation of a conical singularity, that can be obtained by
requiring the proper periodicity for the  Euclidean time in the
Euclidean section of the metric's static regions. In the
near-horizon limit the Euclidean metric for the static regions
takes the form:
\be{ds^2_E\ =\ dR^2
+ \kappa_0^2 R^2 d\tau^2\ + \cdots}\ee
where $R^2= (2p-r)/2p$ and $\kappa_0 = {1}/({4p})$ denotes the
surface gravity at the horizon, as in the previous section.
Demanding no conical singularity at the horizon ($R=0$) requires
the Euclidean time coordinate $\tau$ to be periodic $\tau\sim \tau
+2\pi/\kappa_0$, leading to the Hawking temperature:
\be{ T_H\ =\
\frac{\kappa_0}{2\pi}\ =\ \frac{1}{8\pi p}}.\ee
Ref.~\cite{bqrtz} argued this temperature to be interpretable as
the temperature of the particle distribution seen by static
particle detectors in the static regions.

We now compute a logically unrelated quantity: the particle
production of a massless Klein-Gordon field caused by the
time-dependent fields in region I. To this aim we consider the
massless Klein-Gordon equation\footnote{See \cite{stability} for the
analysis of the black hole case.}:
\be{\frac{1}{\sqrt{-g}}\pl_\mu\left(\sqrt{-g}g^{\mu\nu}\pl_\nu\right)
\Phi \
 = \ 0 ,}\ee
which we again solve by separating variables to obtain the
following mode functions: $u_{m\sigma k}(t,r,\theta,\phi) =
e^{i(m\phi+\sigma r)}\ \Theta_{km}(\theta)\ \Omega(t)+c.c.$. Here
the integer $m$ and real quantities $\sigma$ and $k$ are the
quantum numbers associated with the $\phi$, $\theta$ and $r$
coordinates, respectively. We obtain the following $\theta$
dependence:
\be{{\Theta}_{km}({\theta})\ =\ a\
Q^{\frac{1}{2}(\sqrt{1+4{k^2}}-1)}_{m}\ (\cosh{\theta})}+b\
P^{\frac{1}{2}( \sqrt{1+4{k^2}}-1)}_{m}\ (\cosh{\theta}) \,, \ee
where $a$ and $b$ are integration constants. $P^{r}_{m}(x)$ and
$Q^{r}_{m}(x)$ are the usual associated Legendre functions. In
what follows we restrict our analysis to the simplest case,
${m={k}=0}$.

The physics of interest lies in the $t$-dependent part, which can
be usefully rewritten as a Schr\"odinger-like equation by
performing the substitution
\be \Omega(t) = F(t) \, \left[ {p \over \sqrt{t \, (t - 2p)} }
\right] \, . \label{ChOfVble} \ee
It is also convenient to perform a change of independent variable
${x=t-2p}$ in order to place the horizon at $x = 0$. With these
choices the $t$-dependent equation takes the form:
\be
   \frac{d^{2}F(x)}{dx^2} -  V(x) \, F(x) = 0
\ee
with
\be
 -V(x) = \left(\frac{{\sigma}^2(x+2p)^{2}}{x^{2}}+\frac{2(x+p)}{x(x+2p)^{2}}
+\frac{2(x+p)p}{x^{2}(x+2p)^{2}}-\frac{1}{x(x+2p)}
-\frac{(x+p)^{2}}{x^{2}(x+2p)^{2}} \right)\,. \nonumber
\ee

In order to solve this equation we replace $V(x)$ with an
approximate potential, $V_{\rm approx}(x)$, which is chosen to
properly reproduce the asymptotic form of $V(x)$ as $x \to
\infty$. (This is similar in spirit to replacing $V(x)$ with one
of its Pad\'e approximants.) For these purposes we choose
\be - V_{\rm approx}(x) =
\left(1+\frac{4p}{x}+\frac{4p^2}{x^{2}}\right)
 \sigma^2\,,
\label{approxV} \ee

\vskip 0.2 cm

\noindent We compare the approximate potential with $V(x)$ in
Figs. 3 and 4, for different values of particle label $\sigma$. As
is clear from these figures, the approximate potential follows
$V(x)$ more closely the larger $\sigma$ is and the further $x$ is
chosen from the horizon ($x = 0$).
Remarkably, even for $\sigma p = 1$ the potentials only deviate by
a few percent right at the horizon, where the fractional deviation
becomes $(V - V_{\rm approx})/V \to 1/(16 p^2 \sigma^2 + 1)$. For
this reason we believe the approximate potential to more
accurately capture the form of the Klein Gordon solutions near the
horizon than would be possible using only an asymptotic expansion
of the solutions in powers of $1/(\sigma p)$.

\FIGURE{
\epsfig{file=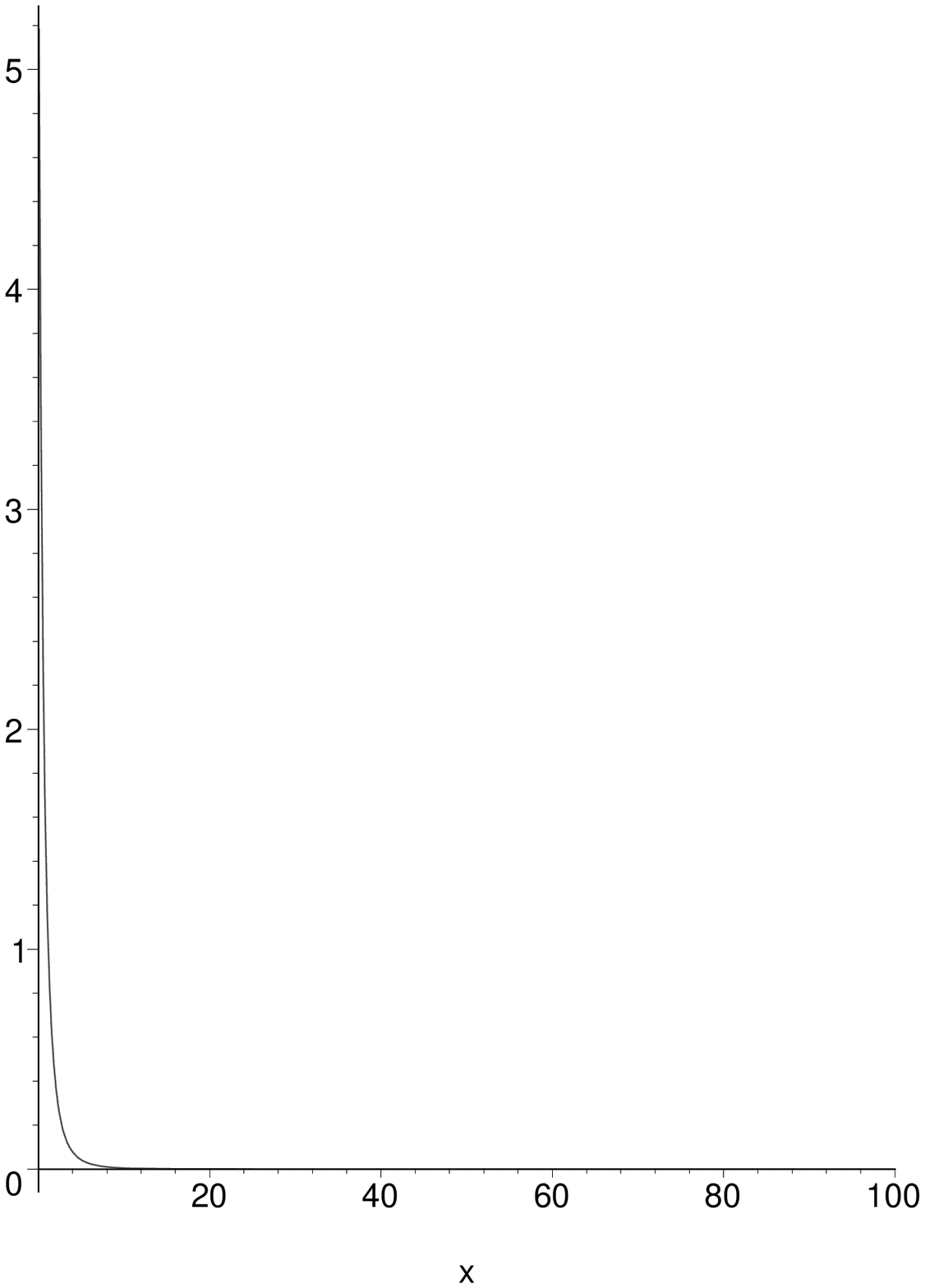, height=2.4in,width=3.8in}
\caption{\it{Percent difference between the function and its
approximation, ${\sigma}p=1$}}}
\FIGURE{ \epsfig{file=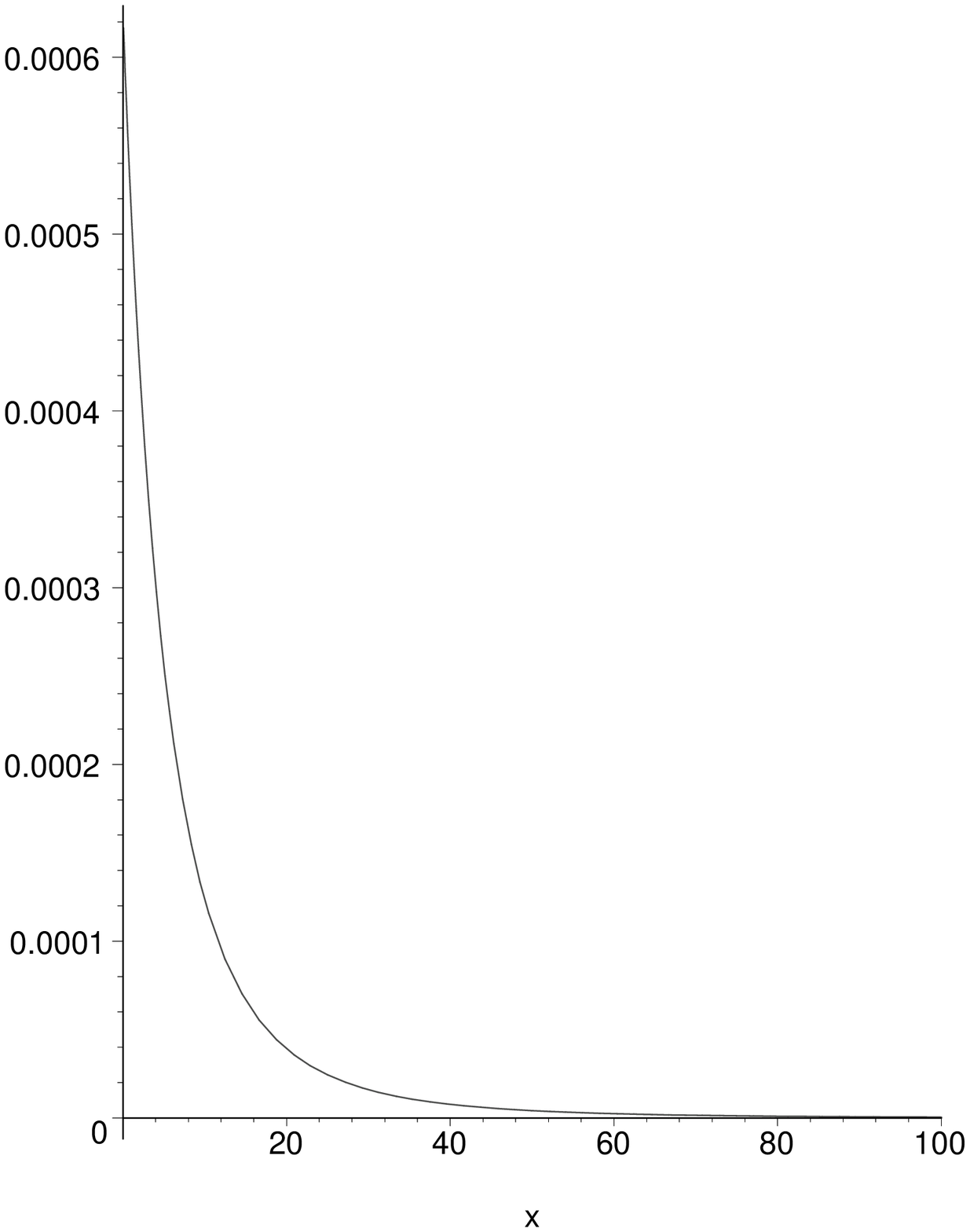,height=2.4in,width=3.8in}
\caption{\it{Percent difference between the function and it's
approximation, ${\sigma}p=100$}}
}

Using the approximate potential the Klein Gordon equation becomes
\be{\frac{d^{2}F(x)}{dx^2}\ +\  \left(1+\frac{4p}{x}+\frac{4p^2}{x^{2}}\right)
 \sigma^2\  F(x)\ =\ 0 \,,
\label{ecuacion}}\ee
which can be solved exactly to give Whittaker functions \cite{GR}
$M_{\chi,\mu}(z)$ and $M_{\chi,-\mu}(z)$, as the
linearly-independent solutions. These are related to standard
confluent hypergeometric functions according to
\be M_{\chi, \mu}(z) = z^{\mu+\frac12} \, e^{-z/2} \; {}_1F_1
\left(\mu- \chi + \frac12, 2 \mu + 1; z \right). \ee
The parameters $\chi$, $\mu$ and $z$ are given in terms of $p$,
$\sigma$ and $x$ by the relations: $\chi = -2i{\sigma}p$,
${z=2i{\sigma}x}$ and $\mu = i \mui$, with $\mui = \frac12 \,
\sqrt{16{\sigma}^{2}p^{2} -1}$. Notice that $\chi$, $\mu$ and $z$
are all pure imaginary so long as $|\sigma p | > \frac14$, as we
shall assume in what follows.

For the purposes of a particle-production calculation we are
interested in the combination of these functions which have
positive and negative frequency near $x = 0$ and $x \to \infty$.
It happens that it is $M_{\chi,-\mu}(z)$ which is positive
frequency near $x = 0$ and $M_{\chi,\mu}(z)$ which is negative
frequency, as may be seen from the small-$z$ limit
\be M_{\chi,\mu}(z) = z^{\mu + \frac12} \, \left[1 + {\cal O}(z)
\right]. \ee
The assignment of positive and negative frequencies follows once
this singular part is re-expressed in terms of the time
coordinate, $x$, in which case $z^{\mu + \frac12} \propto
x^\frac12 \, \exp\left[ i \mui \log(x/p) \right]$. (Recall the
standard phase convention calls $\exp[-i \varphi(t)]$
positive-frequency when $\varphi(t)$ increases with advancing
time, such as for $\varphi = \omega t$ with $\omega > 0$.) For
later purposes we also record here the useful identity:
\be \left[ M_{\chi,-\mu}(z) \right]^* = M_{-\chi,\mu}(e^{-i\pi} z)
= e^{-i\pi(\mu + \frac12)} \; M_{\chi,\mu}(z), \label{ccident} \ee
where we assume $|\sigma p| > \frac14$ in order to use that all
three of $\chi$, $\mu$ and $z$ are pure imaginary.

For $x \to \infty$, on the other hand, it is the particular linear
combination
\be{ W_{\chi,\mu}(z) = \frac{{\Gamma}(-2{\mu})}{{\Gamma}
 (1/2-{\mu}-{\chi})} \; M_{\chi,\mu}(z) + \frac{{\Gamma}(2{\mu})}{{
 \Gamma}(1/2+{\mu}-{\chi})} \; M_{\chi,-\mu}(z)} \,, \label{Wdef} \ee
which is positive frequency, as may be seen from its asymptotic
form
\be W_{\chi,\mu}(z) \sim z^\chi \, e^{-z/2} \; \left[ 1 + {\cal
O}\left({1 \over z} \right) \right], \ee
since $e^{-z/2} = e^{ -i |\sigma| x }$. That it is the absolute
value of $\sigma$ which appears here follows from a careful
treatment of the phase accumulated when $z$ changes sign due to
the branch cut at $z=0$, using identities like eq.~\pref{ccident}.
The additional phase associated with the factor $z^\chi$ does not
change this conclusion. For instance, for $\sigma p \gg 1$ its
effect is simply to change $e^{-i|\sigma|x}$ to
$e^{-i|\sigma|x_*}$, where $x_* = x + 2p \, \log(x/p)$ is the
tortoise coordinate.

With these preliminaries we may now proceed with the
particle-production calculation. If we start with the mode
expansion which is appropriate for large $x$, we have
\be \Phi(x,r,\theta,\phi) = \sum_m \int dk \, d\sigma \; \Bigl[
a_{km\sigma}\ u_{km\sigma}(x,r,\theta,\phi) + c.c. \Bigr] ,
\label{solucion} \ee
where $a_{km\sigma}$ denotes the mode destruction operator and
$u_{km\sigma}(x,r,\theta,\phi) \propto e^{i\sigma r} \,
W_{\chi,\mu}(z)$. 

On the other hand, near the horizon we instead have
\be \Phi(x,r,\theta,\phi) = \sum_m \int dk \, d\sigma \; \Bigl[
b_{km\sigma}\ v_{km\sigma}(x,r,\theta,\phi) + c.c. \Bigr] ,
\label{solucion1} \ee
where $b_{km\sigma}$ are destruction operators and
$v_{km\sigma}(x,r,\theta,\phi) \propto e^{i\sigma r} \,
M_{\chi,-\mu}(z)$. Particle production occurs because the
expansion of $u_{km\sigma}$ in terms of $v_{km\sigma}$ implies
that $b_{km\sigma}$ can be expressed as a linear combination of
$a_{km\sigma}$ and $a^*_{k,-m,-\sigma}$. (It is conservation of
$m$ and $\sigma$ which permits only modes with opposite signs of
$m$ and $\sigma$ to mix in this way.)

For simplicity it is convenient at this point to choose $k=m=0$
and to suppress the $k$ and $m$ labels. The decomposition of
$b_\sigma$ in terms of $a_\sigma$ and $a^*_{-\sigma}$ is found by
choosing a particular $\sigma > 0$ and following those terms whose
$r$-dependence is proportional to $e^{i\sigma r}$. Keeping in mind
that $\chi(-\sigma) = -\chi(\sigma) = -\chi$, $z(-\sigma) = -
z(\sigma) = - z$ and $\mu(-\sigma) = \mu(\sigma) = \mu$, and using
eqs.~\pref{ccident} and \pref{Wdef} we find:
\begin{eqnarray}
    && a_\sigma \, W_{\chi,\mu}(z) + a^*_{-\sigma} \, \Bigl[
    W_{-\chi,\mu}(-z) \Bigr]^* \\
    && \qquad\qquad\qquad =\, {\Gamma(2\mu) \over \Gamma\left(
    \frac12 + \mu - \chi \right)} \left[ a_\sigma  \, M_{\chi,-\mu}(z)
    + a^*_{-\sigma} \, e^{-i \pi \left( \frac12 - \mu \right)} \,
    M_{-\chi,-\mu} \left( e^{-i\pi} z \right) \right] \nonumber\\
    && \qquad \qquad\qquad \sim\, {\Gamma(2\mu) \over \Gamma\left( \frac12
    + \mu - \chi \right)} \;  z^{\frac12 - \mu} \; \Bigl[ a_\sigma +
    a^*_{-\sigma} \, e^{-i \pi \left(1 - 2\mu \right)} \Bigr].
\end{eqnarray}
The last line gives the asymptotic form near $x = 0$.

{}From these manipulations we see that the operators $a_\sigma$
and $b_\sigma$ are related to one another by:
\be b_\sigma = \Lambda(\sigma) \; \Bigl[ a_\sigma - e^{2i\pi \mu} \;
a^*_{-\sigma} \Bigr], \ee
where $\Lambda^2 = 1/[1 - \exp(4i\pi\mu)]$ is determined from the
normalization requirement $[b_\sigma, b^*_\sigma] = [a_\sigma,
a^*_\sigma]$. Inverting this relation gives the expression
\be a_\sigma = \Lambda(\sigma) \; \Bigl[ b_\sigma + e^{2i\pi \mu} \;
b^*_{-\sigma} \Bigr], \ee
which is the main result which is required for the
particle-production calculation.

We imagine preparing the field in the ground state as seen by
observers crossing the horizon: $b_\sigma |0\rangle = 0$, and then
asking for the number of late-time particles which this state
would contain. We find in this way our final result:
\be \label{ppresult+} \langle N_\sigma \rangle \ = {\langle{0}|
a_{\sigma}^{*} \, a_{\sigma}|0\rangle = \frac{1}{e^{4 \pi
\mui}-1}} \,,\ee
where we recall $\mui = \frac12 \, \left(16 \sigma^2 p^2 - 1
\right)^{1/2}$.

This result is very suggestive of a thermal form. Indeed if we
write $\omega = [ \sigma^2 - 1/(4p)^2 ]^{1/2}$, so $\omega \approx
\sigma$ when $\sigma p \gg 1$, then eq.~\pref{ppresult+} is {\it
precisely} thermal,
\be \label{ppresult} \langle N_\sigma \rangle \ =
\frac{1}{e^{8{\pi\omega}p}-1} \, \ee
if $\omega$ is interpreted as the particle energy. (This
interpretation is natural near the horizon where $v_\sigma \sim
z^{-\mu} \sim e^{-i \mui \log x} \sim e^{-2ip \omega \log x} \sim
e^{-i \omega x_*}$ shows that $\omega$ is the eigenvalue of the
operator $i\partial_{t_*}$.) Since our derivation assumes $\mu$ is
pure imaginary it breaks down for $|\sigma p| < \frac14$, where
$\omega$ becomes imaginary. Our approximate form, $V_{\rm
approx}$, also provides a worse description of the full result,
$V$, for $\sigma$ this small. Consequently we cannot yet say
whether these modes are also pair produced at late times.

Even more remarkably, the corresponding temperature is
 \be{T=\frac{1}{8{\pi}p}} \,,\ee
which is exactly the same result obtained earlier by
euclideanizing the metric in the static regions. {\it A priori}
these did not have to agree since the euclidean calculation
describes the particles seen by an accelerating, static observer
behind the horizons, while in the present instance the temperature
corresponds to the distribution of particles which are produced by
the time-dependent fields in region I.

\section{Discussion}
The metrics studied here were proposed in ref.~\cite{bqrtz} with
an eye to using their time dependence for cosmological
applications. Clearly, the classical instability of the horizon
towards singularity formation diminishes the cosmological impact
of these solutions. In particular it prevents the passage from
region III (past, contracting time-dependent solution)
into region I (future, expanding time-dependent solution)
in a way which does not hit a
singularity.
This result is consistent with the strong cosmic censorship
conjecture, since the observer crossing the Cauchy horizon,
decoupling from her past history, would otherwise find a naked
singularity, from which information can come, and yet would be
able to avoid the singularity. It would be interesting to see if
there are cases that avoid this problem \cite{brady,wald}.

A similarly wet blanket is thrown on the S-brane interpretation of
this geometry in terms of a rolling tachyon field
\cite{sbranes,senroll}. In this interpretation we imagine the
rolling of the tachyon field from one minimum of the potential at
$t\rightarrow -\infty $ to the other minimum at $t\rightarrow
\infty$. Each vacuum could then be identified with the asymptotic,
flat infinite past and future of Fig.~1 respectively. The local
maximum of the potential would then be identified with the
horizon. Our result presents an obstruction to this realization of
the rolling by making it impossible to miss a singularity in
between. (This singularity problem is also shared by other
proposed S-brane geometries.)

The existence of the singularity need not invalidate the
interpretation of region I as describing the metric produced by
the late-time rolling of a tachyon from a local maximum to a later
local minimum, however, since this part can be described purely by
the geometry in the future of these singularities. Following this
interpretation, our result for the particle production could be
relevant to the determination of particle production after tachyon
condensation. This would be particularly interesting for the
string scenarios of hybrid inflation from D-brane interactions, as
proposed in \cite{bmqrz}. In this case the particle production
could lead to the determination of the re-heating after inflation
(for a recent discussion of reheating from tachyon condensation
see \cite{cmt}).

We find our particle-production result to be intriguing in its own
right, due to the thermal character of the produced particles, and
the connection which it indicates between the temperature of this
distribution and the temperature obtained by euclideanizing the
metric's static region. This connection is all the more intriguing
given the classical instability which we find, which is likely to
convert the intervening horizons into curvature singularities.

\acknowledgments{
We thank S.-J. Rey for a very  enjoyable collaboration. We  thank
R. Myers, G. Gibbons, D. Easson, F. Gavriil and P.K. Townsend for
 interesting conversations.
We also thank L.~Cornalba and M.~Costa for useful comments on the
first version of the manuscript. 
C.B.'s research is partially funded by grants
from N.S.E.R.C. of Canada and F.C.A.R. of Qu\'ebec. F.Q.'s
research is partially funded by PPARC. G.T. is supported by the European TMR
Networks HPRN-CT-2000-00131, HPRN-CT-2000-00148, and HPRN-CT-2000-00152.
I.Z.C. is supported by CONACyT (Mexico).


\begin{thebibliography}{99} \itemsep -1pt

\bibitem{sbranes}
M.~Gutperle and A.~Strominger,
{\it Spacelike branes,}
 JHEP {\bf 0204} (2002) 018
[hep-th/0202210].



\bibitem{oldsolutions}
D.~L.~Wiltshire,
{\it Global Properties Of Kaluza-Klein Cosmologies,}
Phys.\ Rev.\ D {\bf 36} (1987) 1634;\\
K.~Behrndt and S.~F\"orste,
{\it String Kaluza-Klein cosmology,}
Nucl.\ Phys.\ B {\bf 430} (1994) 441
[hep-th/9403179];\\
A.~Lukas, B.~A.~Ovrut and D.~Waldram,
{\it Cosmological solutions of type II string theory,}
Phys.\ Lett.\ B {\bf 393}, 65 (1997) [hep-th/9608195];\\ {\sl
ibid}
{\it String and M-theory cosmological solutions with Ramond forms,}
Nucl.\ Phys.\ B {\bf 495}, 365 (1997)
[hep-th/9610238];\\
H.~Lu, S.~Mukherji, C.~N.~Pope and K.~W.~Xu,
{\it Cosmological solutions in string theories,}
Phys.\ Rev.\ D {\bf 55} (1997) 7926
[hep-th/9610107];\\
H.~Lu, S.~Mukherji and C.~N.~Pope,
{\it From p-branes to cosmology,}
Int.\ J.\ Mod.\ Phys.\ A {\bf 14} (1999) 4121
[hep-th/9612224].

\bibitem{gqtz}
C.~Grojean, F.~Quevedo, G.~Tasinato and I.~Zavala,
{\it Branes on charged dilatonic backgrounds: Self-tuning, Lorentz  violations and cosmology,}
JHEP {\bf 0108} (2001) 005 [hep-th/0106120].
%

\bibitem{newsolutions}
C.~M.~Chen, D.~V.~Gal'tsov and M.~Gutperle, {\it S-brane Solutions
in Supergravity Theories,} Phys.\ Rev.\ D {\bf 66} (2002) 024043
[hep-th/0204071]; S.~Roy, {\it On
supergravity solutions of space-like Dp-branes,} JHEP {\bf 0208}
(2002) 025 [hep-th/0205198]; A.~Buchel, P.~Langfelder and
J.~Walcher, {\it Does the tachyon matter?,} Annals Phys.\  {\bf
302} (2002) 78 [hep-th/0207235]; N.~Ohta,
{\it Intersection rules for S-branes}, arXiv:hep-th/0301095.
%
\bibitem{rob}
M.~Kruczenski, R.~C.~Myers and A.~W.~Peet, {\it Supergravity
S-branes,} JHEP {\bf 0205} (2002) 039 [hep-th/0204144].
%

\bibitem{bqrtz}
C.~P.~Burgess, F.~Quevedo, S.~J.~Rey, G.~Tasinato and I.~Zavala,
{\it Cosmological spacetimes from negative tension brane backgrounds,}
JHEP {\bf 0210} (2002) 028
[arXiv:hep-th/0207104].\\
F.~Quevedo, G.~Tasinato and I.~Zavala,
{\it S-branes, negative tension branes and cosmology,}
arXiv:hep-th/0211031.

\bibitem{ckk}
L.~Cornalba and M.~S.~Costa,
{\it A New Cosmological Scenario in String Theory,}
Phys.\ Rev.\ D
{\bf 66} (2002) 066001 [hep-th/0203031];
L.~Cornalba, M.~S.~Costa and C.~Kounnas,
{\it A resolution of the cosmological singularity with
orientifolds,}
Nucl.\ Phys.\ B {\bf 637} (2002) 378 [hep-th/0204261].


\bibitem{chandra1}
S.~Chandrasekhar and J.B.~Hartle, Proc.R.Soc. London {\bf A384}, 301 (1982)


\bibitem{chandra}
S. Chandrasekhar, {\it The Mathematical Theory of Black Holes} (Cambridge
University Press, England, 1983)


\bibitem{poisson}
E.~Poisson,
{\it Black-hole interiors and strong cosmic censorship,}
[gr-qc/9709022].


\bibitem{stability}
G.~W.~Gibbons,
{\it An Introduction To Black Hole Thermodynamics,}
LPTENS 80/28.
P.~K.~Townsend,
{\it Black holes,}
arXiv:gr-qc/9707012; \\
J.~Traschen, {\it An introduction to black hole evaporation,}
arXiv:gr-qc/0010055.

\bibitem{brady}
P.~R.~Brady, I.~G.~Moss and R.~C.~Myers,
{\it Cosmic Censorship: As Strong As Ever,}
Phys.\ Rev.\ Lett.\  {\bf 80} (1998) 3432
[gr-qc/9801032].


\bibitem{GR}
 I.S.~Gradshteyn and I.M.~Ryzhik, {\it Table of Integrals Series
 and Products}, Academic Press 1965.

\bibitem{wald}
R.~M.~Wald,
{\it General Relativity,}
{\it  Chicago, Usa: Univ. Pr. (1984) }.

\bibitem{senroll}
A.~Sen,
{\it Rolling tachyon,}
JHEP {\bf 0204} (2002) 048
[hep-th/0203211];\\
A.~Sen,
{\it Tachyon matter,}
 JHEP {\bf 0207} (2002) 065
[hep-th/0203265];\\ A.~Sen,
{\it Field Theory of Tachyon Matter,}
 Mod.\ Phys.\ Lett.\ A {\bf 17}
(2002) 1797 [hep-th/0204143];\\

\bibitem{bmqrz} C.~P.~Burgess, M.~Majumdar, D.~Nolte, F.~Quevedo,
G.~Rajesh and R.~J.~Zhang,
{\it The Inflationary Brane-Antibrane Universe}, JHEP {\bf 0107} (2001) 047,
[hep-th/0105204].

\bibitem{cmt}
G.~Shiu, S.~H.~Tye and I.~Wasserman,
{\it Rolling Tachyon in Brane World Cosmology from Superstring Field Theory,}
[hep-th/0207119]; \\
J.~M.~Cline, H.~Firouzjahi and P.~Martineau,
{\it Reheating from Tachyon Condensation,}, JHEP {\bf 0211} (2002) 041,
[hep-th/0207156].

\end{thebibliography}
\end{document}